\documentstyle[preprint,aps]{revtex}
\begin{document}
\draft

\flushbottom
\preprint{\vbox{
\hbox{IFT-P.xxx/97}
\hbox{hep-yy/9702007}
\hbox{Janeiro 1997}
}}
\title{Collisional Semiclassical Approximations in Phase-Space 
Representation }
\author {G. W. Bund~$^a$\footnote{electronic address: bund@axp.ift.unesp.br} 
S. S. Mizrahi~$^b$\footnote{electronic address: dsmi@power.ufsar.br} and 
M. C. Tijero~$^{a,c}$\footnote{electronic address: maria@axp.ift.unesp.br} }
\address{
$^a$~Instituto de F\'\i sica Te\'orica\\
Universidade Estadual Paulista\\
Rua Pamplona 145\\ 
01405-900-- S\~ao Paulo, SP\\
Brasil}
\address{
$^b$~Departamento de F\'\i sica\\
Universidad Federal de S\~ao Carlos\\
Rodovia Washington Luiz Km 235\\ 
13565-905 S\~ao Carlos, SP\\
Brasil}
\address{
$^c$ Pontif\'\i cia Universidade Cat\'olica (PUC),\\
S\~ao Paulo, SP\\
Brasil}
\maketitle

\newpage
\begin{abstract}
The Gaussian Wave-Packet phase-space representation is used to show that
the expansion in powers of $\hbar$ of the quantum Liouville propagator leads,
in the zeroth order term, to results close to those obtained in the
statistical quasiclassical method of Lee and Scully in the Weyl-Wigner picture.
It is also verified that propagating the Wigner distribution along the 
classical trajectories the amount of error is less than that coming from 
propagating the Gaussian distribution along classical trajectories. 
\end{abstract}

\section{Introduction}
\label{sec:intro}
Nowadays, the advantages and difficulties for using the phase-space 
formulation of quantum mechanics are well known~\cite{ls,hosw,hl}.
This formulation remains still very useful for studying the classical 
limit of quantum mechanics as well as for describing semiclassical
approximations in collisional processes; for these purposes many authors 
use the Weyl-Wigner (WW)picture~\cite{ls,gb,cz,he}. For the collisional
problem it is often combined with the quasiclassical method of 
Ref.~\cite{db}. Following this approach Lee and Scully~\cite{ls} have
improved the accuracy of this method successfully with their Statistical
Quasiclassical (SQC) method which was first suggested by Heller~\cite{eh}.
As we have shown in a previous paper~\cite{bmt}, the approach of Lee and Scully
corresponds to the zeroth order term of the expansion of the quantum 
Liouvillian in powers of $\hbar$ in the WW picture. So, their calculated
transition probabilities could have higher order corrections.

The aim of this article is to show that the use of the Gaussian Wave Packet
(GWP) phase-space representation~\cite{hl,sm,ssm} gives for the zeroth order
term, which corresponds to what we call the Causal Approximation (CA),
results similar to those obtained in the WW representation~\cite{ls}.

In the derivation of the expansion of the quantum Liouvillian in the GWP
picture all orders of $\hbar$ are included; the first order term we shall call
the Quasicausal Approximation (QCA). As it is shown in Refs.~\cite{sm,ssm}
one of the advantages of the GWP representation is that the 
quantum fluctuations of the mapped physical quantities become more evident;
also the distribution function obtained when the density operator for a pure 
state is mapped into this representation is always non-negative.

Here we are
also interested in verifying numerically, for the collisional problem,
the statement put forward by Lee~\cite{hl}: ``the amount of error arising 
from propagating the Wigner distribution function (WDF) along the classical
trajectories is usually considerably less than that coming from propagating 
other distributions along classical trajectories''. In our case the other 
distribution is the GWP distribution~\cite{sm,ssm,mct}. It is in agreement
with Lee's statement the result we derive in Sec. II: first order
corrections in $\hbar$ in the SQC method do not improve the numerical 
results of Lee and Scully~\cite{ls}.

In our comparison of the Weyl-Wigner formalism with the GWP approach we are
going to work with a collinear non-reactive collision of an atom with a 
diatomic molecule, the interaction between them being an exponential 
repulsion, first used by Secrest and Johnson~\cite{sj}, of the form
$V=V_0\exp[-\alpha (x-y)]$ where the constant $V_0$ is related to the 
classical turning point of the trajectory of the particle but its value
does not have any effect on the results. In the WW and GWP formalisms
we have taken this potential, the definition of the $x$ and $y$ coordinates, 
as well as the value of $\alpha$ and $V_0$ from Ref.~\cite{sj}, the parameter
$\alpha$ is adjusted by published experimental data~\cite{sj}. The mapped 
Hamiltonian in the GWP formalism has the constant  $V_0$ renormalized.

In Sec. II we shall summarize our previous paper (Ref.~\cite{bmt}). In Sec. III
we introduce the GWP representation and the CA and QCA. In Sec. IV we 
derive expressions for the transition probabilities and numerical results
are presented in Sec. V. 

\section{Causal and quasicausal approximations in the WW formalism}
\label{sec:ww}
In the first part of this section we are going to give a review of 
Ref.~\cite{bmt}. The quantum Liouville equation in the WW picture is~\cite{gs}
\begin{equation}
\frac{\partial W(q,p,t)}{\partial t}=-i{\cal L}_QW(q,p,t),
\label{11}
\end{equation}
where $(q,p)$ is a point of phase-space and $W(q,p,t)$ is the Wigner 
distribution function (WDF). We are using just one dimension. The quantum 
Liouvillian is
\begin{equation}
{\cal L}_Q=H(q,p)\,\left[ i\frac{2}{\hbar}\,\sin\frac{\hbar}{2}
\stackrel{\leftrightarrow}{\Lambda}\right],
\label{12}
\end{equation}
$H(q,p)$ being the Hamiltonian of the system and the operator
\begin{equation}
\stackrel{\leftrightarrow}{\Lambda} =
\frac{\stackrel{\leftarrow}{\partial}}{\partial q}
\frac{\stackrel{\rightarrow}{\partial}}{\partial p}-
\frac{\stackrel{\leftarrow}{\partial}}{\partial p}
\frac{\stackrel{\rightarrow}{\partial}}{\partial q}
\label{13}
\end{equation}
is the Poisson bracket, arrows indicate on which side the derivatives operate.

In Ref.~\cite{bmt} we show that from the formal solution of Eq.~(\ref{11}) 
given by
\begin{equation}
W(q,p,t)= e^{-i{\cal L}_Q(t-t_0)}W_0(q,p),
\label{14}
\end{equation}
where $W_0(q,p)$ is the WDF at the initial time $t_0$. Taking the classical
limit of ${\cal L}_Q$ in Eq.(\ref{12}) we get
\begin{equation}
    {\cal L}_{cl}=iH(q,p)\stackrel{\leftrightarrow}{\Lambda},
\label{15}
\end{equation}
and Eq.~(\ref{14}) becomes
\begin{equation}
 W^{(0)}(q,p,t)=e^{-i{\cal L}_{cl}(t-t_0)}W_0(q,p)=W_0 
 \left(q(t_0-t),p(t_0-t) \right).
\label{16}
\end{equation}
Thus, each point $(q,p)$ of the phase space of the initial WDF evolves 
classically according to Hamilton's equations, following a classical
trajectory reversed in time. This we call the Causal Approximation (CA).

Still, according to Ref.~\cite{bmt}, we can make an expansion of ${\cal L}_Q$ 
in a power series of $\hbar^2$ which is substituted into Eq.~(\ref{11}) giving
\begin{equation}
    \frac{\partial W(q,p,t)}{\partial t}+i{\cal L}_0W(q,p,t)=
    -i\sum_{n=1}^\infty\hbar^{2n}{\cal L}_{2n}W(q,p,t),
\label{17}
\end{equation}
where ${\cal L}_0={\cal L}_{cl}$ and
\begin{equation}
{\cal L}_{2n}=H(q,p)\left[i\,\frac{(-1)^n}{2^{2n}(2n+1)!} 
\left(\stackrel{\leftrightarrow}{\Lambda} \right)^{2n+1}\right].
\label{18}
\end{equation}
The integral equation corresponding to Eq.(\ref{17}) is given by
\begin{equation}
W(q,p,t)=e^{-i{\cal L}_0(t-t_0)}W_0(q,p)-i\sum_{n=1}^\infty \hbar^{2n}
\int^t_{t_0}dt'e^{-i{\cal L}_0(t-t')}{\cal L}_{2n}W(q,p,t').
\label{19}
\end{equation}
Solving this equation iteratively, we get in first order the Quasicausal
Approximation (QCA)
\begin{equation}
W_{QCA}(q,p,t)=e^{-i{\cal L}_0(t-t_0)}W_0(q,p)-i\hbar^2\int^t_{t_0}dt'
e^{-i{\cal L}_0(t-t')}{\cal L}_{2}e^{-i{\cal L}_0(t'-t_0)}W_0(q,p).
\label{110}
\end{equation}

In Eq.(\ref{110}) the operator $\exp[-i{\cal L}_0(t-t_0)]$ is responsible 
for the classical character of the evolution between different times.

This formalism is applied in Ref.~\cite{bmt} to a collisional process
where a molecule suffers a collision from a pointlike projectile.
As a result the molecule is transferred from the initial discrete energy level
$\vert i\rangle$ to the final level $\vert f\rangle$, the total probability 
for this transition, in the limit $t\to\infty$ and $t_0\to -\infty$,
is given by 
\begin{equation}
{\cal P}_{i\to f}=2\pi\hbar\int^\infty_{-\infty}dp_0\int^\infty_{-\infty}dq_0
W_i(q_0,p_0)e^{i{\cal L}_Q(t-t_0)}W_f(q_0,p_0).
\label{111}
\end{equation}
Eq.(\ref{111}) is an exact result. Introducing now the QCA we get in the limit
$t\to\infty$ and $t_0\to-\infty$
\begin{eqnarray}
{\cal P}^{QCA}_{i\to f}&=&2\pi\hbar\left[\int^\infty_{-\infty}dp_0\int^\infty
_{-\infty}dq_0W_i(q_0,p_0)W_f(q(t-t_0),p(t-t_0))\right.\nonumber \\ 
& + &\mbox{}
\left.i\hbar^2\int^\infty_{-\infty}dp_0\int^\infty_{-\infty}dq_0W_i(q_0,p_0)
\int^t_{t_0}dt'e^{i{\cal L}_0(t-t')}{\cal L}_2W_f(q(t'-t_0),p(t'-t_0))\right],
\label{112}
\end{eqnarray}
where the first term corresponds to the CA, while the second one is the QCA.
The CA corresponds to the statistical quasiclassical (SQC)
method of Lee and Scully given in Ref.~\cite{ls}. Here as in Ref.~\cite{ls}
the $H_2-He$ collision is considered, $H_2$ and $He$ being treated as 
an harmonic oscillator and a free particle respectively. The Hamiltonian 
in the Weyl-Wigner phase space is given by
\begin{equation}
H(Q,q,P,p)=\frac{P^2}{2M}+\frac{p^2}{2m}+\frac{1}{2}\,
k q^2+V_0e^{-\alpha(Q-q)},
\label{113}
\end{equation}
where $Q$ and $q$ are the translational and vibrational coordinates 
respectively, $P$ and $p$ being their respective momenta. All the 
parameters appearing in Eq.(\ref{113}), $M,m,V_0,\alpha$ and $k=m\omega^2$ (elastic 
constant of the oscillator) are taken from Ref.~\cite{sj}. 

Following Lee and Scully (Refs.~\cite{ls,hl}) $Q,q,P,p$ obey 
Hamilton's equations, so they describe classical trajectories, the initial 
state $W_i(q_0,p_0)$ in Eq.(\ref{112}), is given by the WDF for the
harmonic oscillator, the pair $(q_0,p_0)$ refers to the initial position 
and momentum of the harmonic oscillator and it belongs to a two-dimensional 
rectangular grid whose size and density depend on the desired accuracy.

Integrating numerically Hamilton's equations for each $(q_0,p_0)$ of the
grid for the harmonic oscillator and the appropriate initial $Q$ and $P$
of the particle, we get the set of final pairs $(q,p)$ for the oscillator
and final $(Q,P)$ for the particle, initial and final $Q$ must be taken
sufficiently large so that the particle can be considered free, which can
be verified by using the fact that the total energy must be conserved
along the trajectories.

In Eq.(\ref{112}) the final state of the system is now given by the WDF,
$W_f(q,p)$ calculated for all final phase space points of the grid.
Once initial and final WDF are calculated for each point, 
the CA can be obtained using the first term in Eq.(\ref{112}).

In order to obtain the QCA, the second term in Eq.(\ref{112}) must be 
calculated. This term can be approximated by~\cite{mct}
\begin{equation}
C_{QCA}=BF_{if}\int^t_{t_0}dt'e^{-\alpha \left[Q(t'-t_0)-q(t'-t_0)\right]},
\label{114}
\end{equation}
with the constant $B=\pi(\hbar\alpha)^3V_0/12$ and
\begin{equation}
F_{if}=\int^\infty_{-\infty}dp_0\int^\infty_{-\infty}dq_0W_i(q_0,p_0)
e^{i(t-t_0){\cal L}_0}\frac{\partial^3}{\partial p_0^3}\,W_f(q_0,p_0).
\label{115}
\end{equation}

Making $y=2E_{OH}/\hbar \omega$ ($E_{OH}(q_0,p_0)$ being the energy of
the classical harmonic oscillator) we have
\begin{equation}
\frac{\partial^3}{\partial p_0^3}\,W_f(q_0,p_0)=A_0p_0^3+B_0p_0,
\label{116}
\end{equation}
where 
$$
A_0=\left(\frac{2}{\hbar \omega m} \right)^3\,\frac{\partial^3 w_f(y)}
{\partial y^3},\quad\mbox{and}\quad 
B_0=3\left(\frac{2}{\hbar \omega m} \right)^2\,\frac{\partial^2 w_f(y)}
{\partial y^2},
$$
being $w_f(y)=W_f(q_0,p_0)$.

Now we are going to show that this correction, given by Eq.(\ref{114}) 
oscillates periodically in the time t.

Let $t_M$ be the time at which the atom is considered to be a free particle
after colliding with the molecule. At this time the coordinate and momentum
of the classical harmonic oscillator are given by
$$
q_{0M}=A\cos(\omega t_M+\phi_0)\quad \mbox{and}\quad  p_{0M}=-m\omega 
A\sin(\omega t_M+\phi_0).
$$
At a later time $t=t_M+\Delta t$, with $\Delta t>0$, the momentum $p_0(t)$ 
of the harmonic oscillator will be
\begin{equation}
p_0(t)=p_{0M}\cos(\omega \,\Delta t)-m\omega q_{0M}\sin(\omega\, \Delta t).
\label{117}
\end{equation}
Thus, making $a=p_{0M}$ and $b=-m\omega q_{0M}$, one has
\begin{equation}
p_0^3(t)=a^3\cos^3\omega \Delta t+b^3\sin^3\omega \Delta t+3ab^2\cos\omega \Delta
 t\sin^2\omega \Delta t
+3a^2b\sin\omega \Delta t\cos^2\omega \Delta t.
\label{118}
\end{equation}
From Eq.(\ref{117}) and Eq.(\ref{118}) we obtain for Eq.(\ref{116})
\begin{equation}
A_0p^3_0(t)+B_0p_0(t)=
a_3\cos3\omega \Delta t+b_3\sin3\omega \Delta t+a_1\cos\omega \Delta t
+b_1\sin\omega \Delta t,
\label{119}
\end{equation}
where
$$
a_3=A_0\left(\alpha^3+\alpha^{*3}\right),\quad b_3=iA_0(\alpha^3-\alpha^{*3}),
$$
$$
a_1=(3A_0\alpha\alpha^*+B_0)(\alpha+\alpha^*),\quad
b_1=i(3A_0\alpha\alpha^*+B_0)(\alpha-\alpha^*),
$$
and $\alpha=(1/2)(a-ib)$. Eq.(\ref{119}) is a Fourier's series, which 
substituted into Eq.(\ref{115}) shows that $C_{QCA}$ in Eq.(\ref{114}) is 
a periodical function, since in this equation the integral in
the time converges in the limit $t\to\infty$. The average of this periodical
function over one period of the oscillator will be zero. In this derivation
we have used the approximate expression for the correction of the 
transition rate given by Eq.(\ref{114}). The exact demonstration,
although more envolved, follows along similar lines.

\section{Causal and quasicausal approximations in the GWP formalism}
\label{sec:2}
In the GWP representation~\cite{sm,ssm,mct}, operators can be mapped
both into a covariant ($CV$) and a contravariant ($CTV$) form and there are
expressions which relate the $CV$ with the $CTV$ form as well as both
of them with the corresponding $WW$ representation.

The commutator of two operators of the Hilbert space
$A$ and $B$ in the $CV$ form is written~\cite{hl,sm,ssm}

\begin{equation}
\langle pq\vert[A,B]\vert pq\rangle={\cal A}^{CV}(q,p)
\left(
\stackrel{\leftrightarrow}{\Gamma}-\stackrel{\leftrightarrow}{\Gamma ^* }
\right)
{\cal B}^{CV}(q,p),
\label{31}
\end{equation}
where $(q,p)$ is a point in a phase-space, $\vert pq\rangle$ represents 
the minimum uncertainty gaussian wave-packet or coherent state, 
${\cal A}^{CV}(q,p)$ and ${\cal B}^{CV}(q,p)$ are the $CV$ forms of 
operators $A$ and $B$ and  $ \stackrel{\leftrightarrow}{\Gamma}=
exp[{(\hbar/2)\stackrel{\leftarrow}{D}}\stackrel{\rightarrow}{D^*}]$,
where $D=(1/a_0)\partial/\partial q-ia_0\partial/\partial p$. Arrows 
indicate on which side operators act and $a_0$ is a constant with dimensions
$M^{\frac{1}{2}}T^{-\frac{1}{2}}$.

The quantum Liouville equation in the $CV$ representation for the density
operator $P^{CV}(q,p,t)=\langle pq\vert\Psi(t)\rangle\langle\Psi(t)\vert 
 pq\rangle $ is given by
\begin{equation}
\frac{\partial}{\partial t}P^{CV}(q,p,t)=-i{\cal L}^{CV}P^{CV}(q,p,t),
\label{32}
\end{equation}
$ {\cal L}^{CV}$ being the quantum Liouvillian in the $CV$ form
\begin{equation}
{\cal L}^{CV}=\frac{1}{\hbar}{\cal H}^{CV}(q,p)
\left(\stackrel{\leftrightarrow}{\Gamma}-\stackrel{\leftrightarrow}
{\Gamma ^*} \right),
\label{33}
\end{equation}
and ${\cal H}^{CV}(q,p)=\langle pq\vert H\vert pq \rangle $ the $CV$ form
of the hamiltonian $H$~\cite{ssm}.

Now expanding ${\cal L}^{CV}$ in a $\hbar$ power series,
${\cal L}^{CV}=\sum_{n=0}^\infty \hbar^n{\cal L}_n$, we identify
\begin{equation}
{\cal L}_n=\frac{i}{2^n(n+1)!}\,Im\,\left[\left(D^{n+1} 
{\cal H}^{CV} \right) \left(\stackrel{\rightarrow}{D^*} \right)^{n+1} \right],
\label{34}
\end{equation}
where ${\cal L}_0={\cal L}_{cl}$ is the classical Liouvillian.

One defines a Green's function~\cite{ssm} by
\begin{equation}
w(q,p,t\vert q_0,p_0,t_0)=e^{-i(t-t_0){\cal L}^{CV}}w(q,p,t_0
\vert q_0,p_0,t_0),
\label{35}
\end{equation}
with the condition
\begin{equation}
\lim_{t\to t_0}w(q,p,t\vert q_0,p_0,t_0)=\delta(q-q_0)
\delta(p-p_0).
\label{36}
\end{equation}
This Green's function satisfies the Liouville equation
\begin{equation}
\frac{\partial}{\partial t}w(q,p,t\vert q_0,p_0,t_0)=-i
{\cal L}^{CV}w(q,p,t\vert q_0,p_0,t_0),\quad t>t_0
\label{37}
\end{equation}
where the pairs $(q_0,p_0)$ , and $(q,p)$ are the momenta and the 
coordinates at times $t_0$ and $t$ respectively.
Here, as discussed in Ref.~\cite{bmt}, classical causality is broken in the $\hbar$
power series expansion of ${\cal L}^{CV}$ when terms with $n\geq1$ are 
retained.

Because of Eq.~(\ref{34}), Eq.~(\ref{37}) can be written as follows
\begin{equation}
\frac{\partial}{\partial t}w(t\vert t_0)+i{\cal L}_0 w(t\vert t_0)=
\frac{2}{\hbar}\sum^{\infty}_{n=1}\frac{(\frac{\hbar}{2})^{n+1}}{(n+1)!}
Im\, \left[\left(D^{n+1}{\cal H}^{CV} \right) \left(D ^* \right)^{n+1} 
w(t\vert t_0)\right].
\label{38}
\end{equation}

In the GWP phase space the $CV$ representation of the density operator,
${P}^{CV}(q,p,t)$ is given by~\cite{ssm}
\begin{equation}
 {P}^{CV}(q,p,t)=\int^\infty_{-\infty} dp_0
 \int^\infty_{-\infty}dq_0 w(q,p,t\vert q_0,p_0,t_0)
 {P}^{CV}(q_0,p_0,t_0),
\label{39}
\end{equation}
so, if a formal solution of Eq.~(\ref{38}) 
\begin{equation}
w(t\vert t_0)= e^{-i(t-t_0){\cal L}_0}w_(t_0\vert t_0)+\frac{2}{\hbar}\,
\int^{t}_{t_0}dt'e^{-i(t-t'){\cal L}_0}\sum_{n=1}^\infty
\frac{\left(\frac{\hbar}{2} \right)^{n+1}}{(n+1)!}
Im\, \left[\left(D^{n+1}{\cal H}^{CV} \right) \left(D^* \right)^{n+1}
 w(t'\vert t_0)\right]
\label{310}
\end{equation}
is substituted into Eq.~(\ref{39}),
keeping in mind that $w(t_0\vert t_0)=\delta(q-q_0)\delta(p-p_0)$, the
first term gives for ${P}^{CV}(q,p,t)$ the causal
approximation (CA), while  by including the lowest correction $n=1$ term, the 
quasicausal approximation (QCA) is obtained.

\section{Transition probabilities for a collisional process}
\label{sec:sec4}
When we have a system in a given initial state $\vert i\rangle$, at time
$t_0$ and final state $\vert f\rangle$ at time $t$, with $t>t_0$, the
transition probability ${\cal P}_{i\to f}$ may be written~\cite{ssm}
\begin{eqnarray}
 {\cal P}_{i\to f}&=&\frac{1}{2\pi \hbar}\int^\infty_{-\infty}dp
 \int^\infty_{-\infty}dq\,{P}^{CTV}_{i}(q,p,t_0)\,
 e^{i(t-t_0){\cal L}^{CV} } {P}^{CV}_{f}(q,p,t_0)\nonumber \\ &=&
 \frac{1}{2\pi \hbar}\int^\infty_{-\infty}dp \int^\infty_{-\infty}dq
 \,P^{CTV}_{i}(q,p,t_0)\,\tilde P^{CV}_f (q,p,t)
 \label{41}
\end{eqnarray}
being ${P}^{CTV}_{i}$ the CTV distribution function of the initial 
state and ${P}^{CV}_{f}$ the CV distribution function of the final 
state.

For the CTV distribution function we have an equation~\cite{ssm} analogous to 
Eq.~(\ref{32}) whose formal solution is given by
$$
{P}^{CTV}_{i}(q,p,t)=exp\left[-i(t-t_0){\cal L}^{CTV}\right] 
{P}^{CTV}_{i}(q,p,t_0)
$$
with
$$
{\cal L}^{CTV}=\frac{1}{\hbar}{\cal H}^{CTV} \left[
\stackrel{\leftrightarrow}{\Gamma}^{CTV}-\left(
\stackrel{\leftrightarrow}{\Gamma}^{CTV}\right)^*\right],
$$
and
$$
\stackrel{\leftrightarrow}{\Gamma}^{CTV}=exp\left(-\hbar
\stackrel{\leftarrow}{D}^*
\stackrel{\rightarrow}{D}/2\right).
$$

We use the model and method described by Lee and Scully~\cite{ls,hl} for 
the one-dimensional atom-molecule collinear collision, but in the GWP
phase-space formulation. Here also in this non-reactive process the atom
is treated like a free classical structureless particle while the 
molecule is represented by a harmonic oscillator. The interaction
between the atom and the molecule~\cite{sj} is the exponential repulsion
described in Sec. I. The Hamiltonian for this system in 
the $CV$ representation is
\begin{equation}
{\cal H}^{CV}=\frac{P^2}{2M}+\frac{p^2}{2m}+
\frac{1}{2}k q^2+V_{eff}e^{-\alpha(Q-q)}+
\frac{\hbar}{4}\left(\frac{a_0^2}{m}+\frac{k}{a_0^2} \right),
\label{42}
\end{equation}
where $Q$ and $q$ are the translational and vibrational coordinates
respectively, $P$ and $p$ their respective momenta, $V_{eff}=V_0
exp(\alpha^2\hbar/4a_0^2)$ and the parameters $M,m,V_0,\alpha$ and
$k$ (elastic constant) are taken from Ref.~\cite{sj}, fitting the
$He$-$H_2$ system. The $CTV$ form of the Hamiltonian is obtained from 
Eq.~(\ref{42}) by replacing $a_0^2$ by $-a^2_0$~\cite{ssm}.

In the $\hbar$ power-series expansion of ${\cal L}^{CV}$ 
\begin{equation}
{\cal L}^{CV}={\cal L}_0
+\hbar{\cal L}_1+\hbar^2{\cal L}_2+\cdots
\label{43}
\end{equation}
the contributions of the coordinates $Q$ and $P$ were neglected,
except in the zeroth order term given by
\begin{equation}
{\cal L}_0=i\left(
\frac{\partial {\cal H}^{CV}}{\partial Q}
\frac{\partial}{\partial P}-\frac{\partial {\cal H}^{CV}}{\partial P} 
\frac{\partial}{\partial Q}+ 
\frac{\partial {\cal H}^{CV}}{\partial q}
\frac{\partial}{\partial p}-\frac{\partial {\cal H}^{CV}}{\partial p} 
\frac{\partial}{\partial q} 
\right)
\label{44}
\end{equation}
and which corresponds to classical motion.

In this approximation $D^{n+1}{\cal H}^{CV}=(\alpha/a_0)^{n+1}V_{eff}
exp[-\alpha(Q-q)]$ for $n\geq1$, if one takes into account
that $a^2_0=m\omega$.    
Eq.~(\ref{310}) can then be written
\begin{equation}
w(t\vert t_0)=e^{-i(t-t_0){\cal L}_0}w_0(t_0\vert t_0)
+\frac{2 V_{eff}}{\hbar}\int^t_{t_0}dt'e^{-i(t-t'){\cal L}_0}
Im\,\left[ e^{-\alpha(Q-q)}\sum^\infty_{n=2}
\frac{(\frac{\hbar\alpha}{2a_0}D^*)^{n}}{n!}w(t'\vert t_0)\right].
\label{45}
\end{equation}
Substituting Eq.~(\ref{45}) into Eq.~(\ref{39}) we have in first iteration
\begin{eqnarray}
 {P}^{CV}(q,p,t)&=&e^{-i(t-t_0){\cal L}_0}{P}^{CV}(q,p,t_0)
 \nonumber \\ &+&\frac{2V_{eff}}{\hbar}\int^t_{t_0}dt'
 \left[e^{-i(t-t'){\cal L}_0}
 e^{-\alpha(Q-q)}\right]e^{-i(t-t'){\cal L}_0}Im\sum^\infty_{n=2}
\frac{(\frac{\hbar\alpha}{2a_0}D^*)^{n}}{n!}e^{-i(t'-t_0){\cal L}_0} 
{P}^{CV}(q,p,t_0).
\label{46}
\end{eqnarray}

In order to obtain $\tilde P^{CV}(q,p,t)=exp[i(t-t_0){\cal L}^{CV}]
{P}^{CV}_f(q,p,t_0)$, which is the function appearing in Eq.~(\ref{41}),
we proceed in the same fashion, but now as the Green's function is given
by $\tilde w(t\vert t_0)=exp[i(t-t_0){\cal L}^{CV}] w(t_0\vert t_0)$,
we get instead of Eq.~(\ref{46}) 
\begin{eqnarray}
\tilde{{P}}^{CV}(q,p,t)&=&e^{i(t-t_0){\cal L}_0}{P}^{CV}(q,p,t_0)
 \nonumber \\ &-&\frac{2V_{eff}}{\hbar}\int^t_{t_0}dt'
 \left[e^{i(t-t'){\cal L}_0}
 e^{-\alpha(Q-q)}\right]e^{i(t-t'){\cal L}_0}Im\sum^\infty_{n=2}
\frac{(\frac{\hbar\alpha}{2a_0}D^*)^{n}}{n!}e^{i(t'-t_0){\cal L}_0} 
{P}^{CV}(q,p,t_0)
\label{47}
\end{eqnarray}
since $\tilde P^{CV}(q,p,t_0)=P^{CV}(q,p,t_0)$.

Substituting Eq.~(\ref{47}) into Eq.~(\ref{41}) we have, with quantum
corrections in all orders of $\hbar$
\begin{eqnarray}
{\cal P}_{i\to f}&=&\int\frac{dpdq}{2\pi\hbar}{P}^{CTV}_i(q,p,t_0)
e^{i(t-t_0){\cal L}_0}{P}^{CV}_f(q,p,t_0)\nonumber  \\  & &\mbox{}
-\frac{\alpha V_{eff}}{a_0}\int\frac{dpdq}{2\pi\hbar}
{P}^{CTV}_i(q,p,t_0)\int^t_{t_0}dt'\left[
e^{i(t-t'){\cal L}_0}e^{-\alpha(Q-q)}\right]\nonumber \\ & &\mbox{}
\cdot e^{i(t-t'){\cal L}_0}\,Im\,\left[ \left(\int^1_0d\xi
e^{\frac{\xi\hbar\alpha}
{a_0}D^*}-1 \right) D^*e^{i(t'-t_0){\cal L}_0}{P}^{CV}_f(q,p,t_0)\right],
\label{48}
\end{eqnarray}
where we have used the property, valid for any operator $A$
\begin{equation}
\int^1_0 d\xi e^{\xi A}=\sum^\infty_{n=0}\frac{A^n}{(n+1)!}.
\label{49}
\end{equation}

The first term in Eq.~(\ref{48}) is the CA which is the zeroth
order term of our $\hbar$ power series-expansion, and corresponds to the 
expression of the Statistical Quasiclassical (SQC) method of Lee and 
Scully~\cite{ls}. The difference between these two expressions lies in the 
distribution functions, while Lee and Scully work with the product of
two Wigner Distribution Functions (WDF), in Eq.~(\ref{48}) we have
the product of two Gaussian Distribution Functions (GDF), one of them
in the $CTV$ form and the other in the $CV$
form.

Like in the SQC method,~\cite{ls,hl} here ${P}^{CTV}_i(q,p,t_0)$ 
represents the initial vibrational state of the system, which 
we propagate along the classical trajectories.
 We also use the Lee and Scully method~\cite{ls} for 
constructing the two-dimensional rectangular grid in the $(q,p)$ plane.
${P}^{CTV}_i(q_n,p_n,t_0=-\infty)$ is the weight carried by the point 
$(q_n,p_n)$ of the $n$th cell of the grid.

After integrating Hamilton's equations for each point of the grid, the final
GDF (${P}^{CTV}_f(q,p,t=\infty)$) is calculated and then the transition
probabilities are computed in the CA which corresponds to the first 
term in Eq.~(\ref{48}).

The GDF in the $CV$ form is given by (see Appendix)
\begin{equation}
{P}^{CV}_n(y)=\frac{1}{n!}\,y^ne^{-y},
\label{410}
\end{equation}
where $y=E/\hbar w$, $E(q,p)$ being the energy of the classical harmonic 
oscillator, $n$ is the quantum number
which corresponds to the $n$th eigenstate of the harmonic
oscillator. Introducing the new variable $r^2=y$, the GDF in the $CTV$ form
is (see Appendix)
\begin{equation}
{P}^{CTV}_n(r)=e^{-\frac{1}{4r}\frac{\partial}{\partial r}
r\frac{\partial}{\partial r}}\,{P}^{CV}_n(r).
\label{411}
\end{equation}

\section{Numerical results and conclusions}
\label{sec:num}
Using the mapped form of the quantum Liouville equation into the GWP
phase-space we have derived expressions for the transition probabilities
for semiclassical calculations of inelastic atom-molecule 
collisions which include also the quantum terms. These expressions 
are similar to those obtained in Ref.~\cite{bmt}
for the WW formalism.

We show in Tables I and II the transition probabilities ${\cal P}_{i\to f}$
from the initial $(i)$ to the final $(f)$ state of the $He$-$H_2$ system
which were computed first by integrating numerically the Schr\"odinger
equation, (these are the exact quantum mechanical
results (QM) taken from Ref.~\cite{sj}), second  by using the Weyl-Wigner
representation in the statistical quasiclassical (SQC) method taken from
Ref.~\cite{ls} and third by using the gaussian wave-packet representation
in the causal approximation (GWP). The last ones are our results and
they are given by the first term in Eq.~(\ref{48}). For more details on
the system and models see the mentioned references.

The results in Tables I and II obtained using only  the zeroth order term
for both methods SQC and GWP should improve if quantum corrections
were introduced.

The accuracy of the SQC and the GWP methods was checked by studying how well
the transition probabilities obey microscopic reversibility, the results
are presented in Table III. These results confirm that for the CA,
the amount of error coming from propagating
the WDF along classical trajectories is less than that coming from propagating
the GWP distribution also along the classical trajectories, as stated
in Ref.~\cite{hl}. This agrees with the vanishing result found 
in Section II for the correction to the CA in the WW formalism. However,
preliminary calculations of the first quantum correction (QCA) in the GWP
formalism gave nonzero results.

Both methods (SQC and GWP) give nonvanishing results for classically
forbidden processes. The numerical calculations, as far as the CA
is concerned, for both formalisms present almost the same degree of
difficulty for being performed.  

\acknowledgments
One of us (M. C. T.) would like to thank the Instituto de F\'\i sica
Te\'orica for the hospitality.

\appendix
\section*{Derivation of the $CTV$ form of the
$GDF$}
\bigskip
Let $\vert \varphi _n \rangle$, with $n=0,1,2...$ be the eigenstates of the
harmonic oscillator and $\vert \alpha \rangle = \vert pq \rangle $ a 
coherent state, the $CV$ form for the density operator~\cite{sm,whl}
$\langle pq \vert \varphi _n \rangle \langle \varphi _n \vert pq \rangle $ 
is given by $P^{CV}_n(r)=\frac{r^{2n}}{n!}e^{-r^2}$, where 
$\alpha=(a_0^2/2\hbar)^{\frac{1}{2}}q+i(1/2\hbar a_0^2)^{\frac{1}{2}}p 
= r e^{i\theta}$.
If $E$ represents the energy of the classical harmonic oscillator and
$r^2=\frac{E}{\hbar \omega} = y$, we may write
\begin{equation}
P^{CV}_n(y)=\frac{1}{n!}y^n e^{-y}.
\label{511}
\end{equation}
We see that this $CV$ form of the $GDF$ depends only on the energy of
the harmonic oscillator.

In order to obtain the $CTV$ form of the $GDF$ we use the property of
the coherent states~\cite{whl} which relates the normal $(NO)$ and
antinormal $(AO)$ ordering of a function $f$ of $\alpha$ and $\alpha ^*$
\begin{equation}
f^{(AO)}(\alpha , \alpha^*)=e^{\frac{\partial^2}{\partial \alpha \partial
\alpha^*}}f^{(NO)}(\alpha ,\alpha^*).
\label{512}
\end{equation}
In polar coordinates $r$ and $\theta$, we have
\begin{equation}
\frac{\partial^2}{\partial \alpha \partial \alpha^*}=
\frac{1}{4 r}\left( \frac{\partial}{\partial r}r \frac{\partial}
{\partial r}+ \frac{1}{r} \frac{\partial^2}{\partial \theta^2} \right),
\label{513}
\end{equation}
and Eq.~(\ref{512}) in the $GWP$ representation is written~\cite{sm}
\begin{equation}
P^{CTV}_n(r,\theta)=e^{-\frac{1}{4r}\left(\frac{\partial}{\partial r}
r \frac{\partial}{\partial r} + \frac{1}{r} \frac{\partial^2}
{\partial \theta^2}\right)} P^{CV}_n(r,\theta).
\label{514}
\end{equation}
For any positive integer $n$ we have derived the formula
\begin{equation}
\left(\frac{1}{4r} \frac{\partial}{\partial r} r \frac{\partial}{\partial r}
\right)^n e^{-r^2}= \left(-1\right)^n n! L_n\left(r^2 \right) e^{-r^2},
\label{515}
\end{equation}
where $L_n$ are the Laguerre polynomials.

Since $P^{CV}_n (r)$ for the harmonic oscillator do not depend on $\theta$,
Eq.~(\ref{514}) can be written as
\begin{equation}
P^{CTV}_n (r)= e^{- \frac{1}{4r} \frac{\partial }{\partial r}
r \frac{\partial}{\partial r}} P^{CV}_n (r).
\label{516}
\end{equation}
Using Eqs.~(\ref{516}), Eq.~(\ref{511}) and Eq.~(\ref{515}) we obtain
for $n=0$
\begin{equation}
P^{CTV}_0(r)=\sum^{\infty}_{n=0} L_n(r^2) e^{-r^2}.
\label{517}
\end{equation}
For $n=1$ we get, similarly,
\begin{equation}
P^{CTV}_1 (r)=\sum^{\infty}_{n=0} (1-n) L_n (r^2) e^{-r^2},
\label{518}
\end{equation}
because
\begin{equation}
\left(\frac{1}{4r} \frac{\partial}{\partial r} r \frac{\partial}
{\partial r}\right)^n \left( r^2 e^{-r^2}\right)=
\left(-1 \right)^{n+1} n! \left[ \left(n+1\right)L_{n+1}(r^2)-
L_n(r^2)\right] e^{-r^2}.
\label{519}
\end{equation}
For $n=2$
\begin{equation}
P^{CTV}_2 (r)= \frac{1}{2!}r^4 e^{-r^2} + \left[\sum^{\infty}_{n=3}
\frac{n(n-1)}{2} L_n(r^2) - 2 \sum^{\infty}_{n=2} n L_n(r^2)
+ \sum^{\infty}_{n=1} L_n (r^2) \right] e^{-r^2}.
\label{520}
\end{equation}

\newpage
\begin{center}
{\bf{\large Table Captions}}
\end{center}
\noindent {\bf Table I.} Transition probability ${\cal P}_{0\to f}$ for a 
collinear He-$H_2$ collision calculated by the quantum mechanical (QM) 
method (Ref.~\cite{sj}), statistical quasiclassical (SQC) method 
(Ref.~\cite{ls}) and gaussian wave-packet (GWP) method in the CA 
(first term in Eq.(\ref{48})).
The total initial energy E is measured in units of $\hbar \omega/2$, where
$\omega$ is the vibrational frequency of the $H_2$ molecule.
In the calculation of ${P}_0^{CTV}$ at least 100 Laguerre polynomials
were used, although only about 10 polynomials are required in order to
obtain convergence. Numbers inside brackets give
the upper limit of the transition probability and $*$ means that
the transition is prohibited classically.
\vskip .5cm
\noindent {\bf Table II.} Similar to Table I; except that the oscillator 
goes from the first exited state to the final state $f$.
\vskip .5cm
\noindent {\bf Table III.} Similar to Table I; results of Table I and
Table II are compared in order to test microscopic reversibility. 

\newpage
\begin{table}
\begin{tabular}{|ccccc|} \hline
E & ${\cal P}_{0\to f}$ &  QM     & SQC   & GWP \\ \hline
8 & $0\to 0$     & (0.892) & 0.893 & 0.830 \\
  & $0\to 1$     & 0.108   & 0.107 & 0.156 \\  
  & $0\to 2$     & 0.001   & -     & 0.014 \\ \hline
12 & $0\to 0$    & (0.538) & 0.529 & 0.501 \\
   & $0\to 1$    & 0.394   & 0.412  & 0.349 \\
   & $0\to 2$    & 0.068   & 0.068  & 0.122 \\
   & $0\to 3$    & -       &   -    & 0.028 \\ \hline
16 & $0\to 0$    & (0.204) & 0.187  & 0.229 \\
   & $0\to 1$    & 0.434   & 0.422  & 0.339 \\ 
    & $0\to 2$   & 0.291   & 0.314  & 0.250 \\
    & $0\to 3$   & 0.071   & 0.077 & 0.124 \\
    & $0\to 4$   & -       & -     & 0.045 \\
    & $0\to 5$   & -       & -     & 0.013 \\ \hline
 20 & $0\to0$    & $(0.060)^*$     &  $ 0.046^*$  & $ 0.090^*$ \\
    & $0\to 1$   & 0.128           & 0.202         & 0.221 \\
    & $0\to2$    &  0.366           & 0.351        & 0.270 \\
    & $0\to3$    & 0.267            & 0.294        & 0.220 \\
    & $0\to 4$   & 0.089            & 0.106        & 0.134 \\
    & $0\to5$    & -                & -            & 0.66  \\  \hline
\end{tabular}
\caption{}
\end{table}
\newpage
\begin{table}
\begin{tabular}{|ccccc|}\hline
E   & ${\cal P}_{1\to f}$    & QM     & SQC   & GWP \\ \hline
8   & $1\to 0$               & 0.108       & 0.106  & 0.135 \\
    & $1\to 1$               & (0.850)     & 0.863 & 0.780 \\
    & $1\to 2$               & 0.042       & 0.031 & 0.085  \\ \hline
12  & $1\to 0$               & 0.394       & 0.411 & 0.396  \\
    & $1\to 1$               & (0.244)     & 0.176 & 0.250  \\
    & $1\to 2$               & 0.345       & 0.385 & 0.272  \\
    & $1\to 3$               & $0.037^*$   & $0.028^*$ & $0.082^*$ \\ \hline
16  & $1\to 0$               & 0.434       & 0.420     & 0.377 \\
    & $1\to 1$               & (0.034)     & 0.065     & 0.137 \\
    & $1\to 2$               & 0.220       & 0.151     & 0.176  \\
    & $1\to 3$               & 0.261       & 0.302     & 0.181  \\
    & $1\to 4$               & 0.051       & 0.061     & 0.098  \\
    & $1\to 5$               &   -         &   -       & 0.031   \\ \hline
20  & $1\to 0$               & 0.218       & 0.199     & 0.230   \\  
    & $1\to 1$               & (0.286)     & 0.285     & 0.233   \\
    & $1\to 2$               & 0.009       & 0.042     & 0.143    \\
    & $1\to 3$               & 0.170       & 0.090     & 0.149    \\
    & $1\to 4$               & 0.240       & 0.262     & 0.149   \\
    & $1\to 5$               & 0.077       & 0.121     & 0.105 \\ \hline
\end{tabular}
\caption{}
\end{table}
\newpage
\begin{table}
\begin{tabular}{|ccccc|} \hline
E    &   ${\cal P}_{i\to f}$   & QM     & SQC     & GWP    \\ \hline
8    &   $0\to 1$              & 0.108  & 0.107   & 0.156   \\
     &   $1\to 0$              & 0.108  & 0.106   & 0.135  \\ \hline
12   &   $0\to 1$              & 0.394  & 0.412   & 0.349  \\
     &   $1\to 0$              & 0.394  & 0.411   & 0.396  \\ \hline
16   &   $0\to 1$              & 0.424  & 0.422   & 0.339  \\
     &   $1\to 0$              & 0.434  & 0.420   & 0.377   \\ \hline
20   &   $0\to 1$              & 0.218  & 0.202   & 0.221   \\
     &   $1\to 0$              & 0.218  & 0.199   & 0.230 \\ \hline
\end{tabular}
\caption{}
\end{table}
\end{document}